\documentclass{aa}

\begin{document}

   \thesaurus{     
              12.12.1;  %
              13.07.1} %
   \title{On the existence of the
    intrinsic anisotropies in the angular distributions of gamma-ray
          bursts}
    \titlerunning{On the existence of the intrinsic anisotropies in the
angular distributions ...}

   \author{ A. M\'esz\'aros  \inst{1}
            \and Z. Bagoly \inst{2}          
            \and R. Vavrek \inst{3,4}}
   \authorrunning{A. M\'esz\'aros et al.}
    
   \offprints{A. M\'esz\'aros}

   \institute{Department of Astronomy, Charles University,
              V Hole\v{s}ovi\v{c}k\'ach 2, CZ-180 00 Prague 8,
              Czech Republic\\
              (meszaros@mbox.cesnet.cz)
          \and Laboratory for Information Technology, E\"{o}tv\"{o}s
               University, H-1117 Budapest, P\'azm\'any P\'eter s\'et\'any
               1./A, Hungary\\
               (bagoly@ludens.elte.hu)
          \and  Konkoly Observatory, Budapest, Box. 67, H-1505 Hungary  
                (vavrek@ogyalla.konkoly.hu)                            
          \and Observatoire de Paris, F-92195, Meudon Cedex, France}        

   \date{Received 5 July, 1999 / Accepted <date>}

  \maketitle

 \begin{abstract}

This article is concerned primarily with the intrinsic anisotropy 
in the angular distribution of 2281 gamma-ray bursts (GRBs)
collected in Current BATSE Gamma-Ray Burst Catalog until the end of year 1998,
and, second, with intrinsic anisotropies 
of three subclasses ("short", "intermediate", "long") of GRBs.
Testing based on spherical harmonics of each class, in equatorial coordinates, 
is presented. Because the sky exposure function of BATSE instrument is not 
dependent on the right ascension $\alpha$, any non-zero
spherical harmonic proportional either to
$P_n^m(\sin \delta) \sin m \alpha$ or to
$P_n^m(\sin \delta) \cos m \alpha$ with $m \neq 0$ 
($\delta$ is the declination),
immediately indicates an intrinsic non-zero term. It is
a somewhat surprising result that the "intermediate" 
subclass shows an intrinsic anisotropy at the $97\%$ significance level
caused by the high non-zero $P_3^1(\sin \delta) \sin \alpha$ harmonic.
The remaining two subclasses, and the full sample of 
GRBs, remain isotropic.

   \end{abstract}

\section{Introduction}

The discovery of an anisotropy in the angular distribution of 
2025 gamma-ray bursts (GRBs) collected in Current BATSE Gamma-Ray 
Burst Catalog (\cite{meegan}) up to the end of year 1997, has recently 
been announced (\cite{balazs98}). 
Of course, the existence
of an observed anisotropy is expected from the non-uniform sky
exposure function of BATSE instrument (\cite{fishman94}). 
Nevertheless, Bal\'azs et al. (1998) argue that the 
detected anisotropy should {\it not} be a pure instrumental 
effect; i.e. there also exists an {\it intrinsic} anisotropy
(see also Bal\'azs et al. 1999).
The exact details of the nature of the 
intrinsic anisotropy remains  open to debate. 

Recently, Horv\'{a}th (1998) and Mukherjee et al. (1998) have proposed 
that the long GRBs should be further separated into two classes; 
the limit separating the "long" subclass should be at 
$T_{90} = 10$ s, where $T_{90}$ is 
the "90\%" duration (Fishman et al. 1994). 
Hence, in the studies of subclasses
it is required to consider
{\it three} subclasses separately ("short" subclass with $T_{90} < 2$ s;
"intermediate" subclass with $2$ s $ < T_{90} < 10$ s; 
"long" subclass with $T_{90} > 10$ s).  

In principle, for GRBs, there are 16 possible types which can occur, 
because for "all" GRBs, and for the three subclasses as well, there are 
two eventualities: either they are distributed isotropically or they
are not.

The purpose of this paper is to 
clarify the situation concerning these possibilities. 
We will again use the analysis based on the spherical 
harmonics (\cite{balazs98}).

In the study by Bal\'azs et al. (1998), an analysis of GRBs having a 
peak flux greater than 0.65 photons/(cm$^2$ s) was performed. Hence,
the GRBs having a peak flux smaller than 0.65 photons/(cm$^2$ s) were omitted
from the sample.
This truncation did not give any new result; except for the fact that the 
number of GRBs was lo\-wered, and hence the corresponding significance 
levels were also lowered. 
The High-Energy versus Non-High-Energy separation
(\cite{pendleton}) did not give any new result either;
hence, we will not use this separation here. 

It is well-known that there is no
general agreement concerning the significance 
level required to reject a null hypothesis.
In an ad hoc manner, we will take, in accordance with the propositions 
of standard statistical textbooks (\cite{trumpler}, 
\cite{kendall}), for the  "doubtless" or "definite" 
rejection, the 
$\geq 99.9\%$ significance level.  
(This level of significance is obtained for all GRBs in the sample
examined by Bal\'azs et al. 1998; 
but together with the instrumental effects.)
We will treat a significance level of $\geq 99\%$ as a 
{\em practically} sure result, with the $\geq 95\%$ significance level 
result being considered as {\em noteworthy}. 

   \section{Mathematical considerations}

The key ideas of the analysis by Bal\'azs et al. (1998) were based on 
spherical harmonics up to the quadrupole terms. In this section, we generalize 
these considerations to include higher order harmonics. In addition, we 
introduce new statistical tests.

As we note in the introduction, it is necessary to eliminate
the BATSE instrumental effects co\-ming from 
the non-uniform sky exposure. Several methods exist. 
For example, in Briggs et al. (1996), Monte Carlo simulations are used:
$N$ points are randomly scattered on the sky many times, and
the scattering is "deformed" in accordance with the sky exposure function; 
$N$ is the number of GRBs. Then the spherical harmonics calculated from 
these simulations are compared with the sphe\-rical harmonics of the observed 
BATSE data. Moreover, in Briggs et al. (1996) a further quite simple method is 
proposed: to use the equatorial coordinates $\alpha$ and $\delta$ 
($0 \leq \alpha \leq 2\pi$; $-\pi/2 \leq \delta \leq \pi/2$) instead 
of the Galactic coordinates, because the sky exposure function is not 
dependent on $\alpha$. 
 In this paper we will use 
these coordinates and the proposed "equatorial" method.

The key ideas of this method may be seen as follows.
 Let $\omega(\alpha, \delta)\; \cos \delta\; d\delta \;
 d\alpha$ be the probability
 of finding a GRB in the infinitesimal solid angle
 $\cos \delta\; d\delta\; d\alpha$. If we assume that there is an
isotropy in the sky distribution of GRBs, then $\omega$ should reflect the
sky exposure function (\cite{balazs98}). The function $\omega$ can be decomposed into its
 spherical harmo\-nics. One then has

   $$
   \omega(\alpha, \delta) = \sum_{n=0}^{\infty} \omega_{n,0}
P_n(\sin \delta) +
   $$
\begin{equation}
    \sum_{n=1}^{\infty}\sum_{m=1}^{n} P_n^m(\sin \delta)
    (\omega_{n,-m} \sin m\alpha +
    \omega_{n,m} \cos m\alpha),
      \end{equation}
where $P_n^m(\sin \delta)$ are the Legendre polynomials.

   If there is an intrinsic isotropy, then
 $\omega(\alpha, \delta) = \omega( \delta)$, and one must have
$\omega_{n, -m} = \omega_{n,m} = 0$ for any $n \geq m \geq 1$.
Conversely, if even one single $\omega_{n, \pm m}$ for $m \geq 1$
were non-zero, then there would be an intrinsic anisotropy. {\it This is the
key idea of this method.} Simply, if $\omega$ is not dependent on
$\alpha$, then all $m \neq 0$ harmonics must be identically zero
independent  
of the concrete form of function $\omega(\delta)$. Then, of course,
this theoretical prediction should be compared to the observational data.

 Concerning the $m = 0$
terms one may use either the Monte Carlo method (\cite{briggs}), or - if
possible - to calculate from the 
known $\omega(\delta)$, numerically or even analytically, the expected 
theoretical values of $\omega_{n,0}$. 
They will not be zeros in the general case.
(They would be zeros, if $\omega$ were constant; i.e. if the sky exposure
function were constant.)

In this paper we will not study
these terms in detail, and an eventual observed
non-zero $\omega_{n, 0}$ will simply be assumed to be that caused 
exclusively by the instrumental effects.

In our case the null hypothesis of intrinsic
isotropy will be identical to the assumption
$\omega(\alpha, \delta) = \omega(\delta)$; i.e. to the assumption that
for all $\omega_{n, \pm m} = 0$ for $m \neq 0$.
Hence, to test the null hypothesis of intrinsic isotropy one has to test
these theoretically expected zeros for the given spherical harmonics with
$m \neq 0$.

This may be done similarly to Bal\'azs et al. (1998). 
This means that, if one has $N$
GRBs with measured $[\alpha_j, \delta_j]$ ($j=1,2,...,N$) coordinates,
one has to calculate for the given $n, m$ the values
(Peebles 1980, Eq. (46.1); Bal\'azs et al. 1998)
\begin{equation}
\sum_{j=1}^{N} P_n^m(\sin \delta_j) \cos m\alpha_j
 \end{equation}
  and
\begin{equation}
\sqrt{ \sum_{j=1}^{N} (P_n^m(\sin \delta_j) \cos m\alpha_j)^2}.
 \end{equation}
(For $n, -m$ one simply has to substitute cosine by
sine.) Then one can apply  Student's $t$ test, where
 the expected mean is zero. Here  $t$ is simply the ratio
 of quantities from Eqs. (2-3), because for $N \gg 1$ (this is always the
case here) one may use $N-1 \simeq N$. Note also that there are 
no problems  which will arise from the normalization constant of
a spherical harmonic. (If,
instead of $P_n^m$, one uses $const. P_n^m$, where $const.$ is an arbitrary
positive number, then  $t$ will not change, if $N-1 \simeq N$,
and if the expected mean is zero.)
Technically, there can also occur several computational 
numerical problems with the
 definition of $P_n^m$ itself for large $n$ on the computer
(see Press et al. 1992, Chapt. 6.8).
In our calculations, up to the used $n$, we encountered no such
problems. We used the algorithm described in Chapt. 6.8 of
Press et al. (1992).

We will also use a modification of this test. From the symmetry properties
of $P_n^m(\sin \delta) \cos m \alpha$ and 
$P_n^m(\sin \delta) \sin m \alpha$ it follows that for $m\neq 0$ in exactly
one half of the sky the sign of the harmonic is positive, 
and in the other  half it is negative.
For $m = 0$ this is true for odd $n$ only, but not for even $n$. Hence,
except for even $n$ with $m=0$, one may proceed as follows. One simply
calculates the number $k_{\mathrm{obs}}$ of GRBs having - say - negative values. Then
one may apply the Bernoulli test described in M\'esz\'aros (1997) and
Bal\'azs et al. (1998). (It is essential to note that
the half of the sky, where the given harmonic has the given sign, 
need not be a "connected" compact region (\cite{balazs98}).) Then
for the given $k_{\mathrm{obs}obs}$ the calculation of 
the significance level is described in Bal\'azs et al. (1998).
In essence, one can take - as the theoretical expectation - a normal
distribution with mean $N/2$ and dispersion $\sqrt{N}/2$. Hence, if
$|k_{\mathrm{obs}} - N/2| > \sqrt{N}$, then the significance level is larger than 
$2\sigma$, i.e. $95.4\%$ (\cite{balazs98}). 
Note that - trivially - in this "sign test"
the choice sign does not matter. Obviously, the "sign-test" and  Student's $t$ test need
not give the same result for a given harmonic.
 
This method based on spherical harmonics
seems to be extremely powerful because, at least in principle,
one single non-zero harmonic for $m \neq 0$ should be enough
to reject the null hypothesis of intrinsic isotropy
(Peebles 1980, Chapt. 46; Tegmark et al. 1996). This follows from the 
orthogonality of spherical functions (Peebles 1980, Chapt. 46).
On the other hand, it is not excluded that - testing several harmonics
simultaneously with two different tests -, 
some of these tests can give a higher than $\alpha \%$ significance
level "by chance". For example, if $\alpha = 95$, then one may expect
that $5\%$ of tests can give a 
greater than $95\%$ significance level "by chance".
Therefore, to be maximally careful, we will reject the null hypothesis 
only in the case, when the number of tests giving $\geq \alpha \%$ 
significance level is far above $(100-\alpha)\%$ fraction. Concretely, this 
means the following. We take the absolute value of the difference between 
the number of tests giving $\geq \alpha \%$ significance level and 
the theoretically expected number of tests giving $\geq \alpha \%$ 
significance level. This value must be so large
that the probability of a chance of this
difference is smaller than $(100 -\alpha)\%$. This probability is
simply calculated from the binomial (Bernoulli) distribution.

In our case we have infinite harmonics and hence in principle we may apply an infinite number of 
tests.
Nevertheless, due
to the discrete character of measured $\omega(\alpha,\delta)$, there is
 a limitation to the harmonics, which can be tested. Having
the angular coordinates of $N$ GRBs, we have $2N$ independent measured
quantities. Hence, clearly, we cannot obtain more than $2N$ {\it independent}
harmonics. Which harmonics
are these? In addition, even for these $2N$ independent
harmonics there
may occur a further problem. Of course, the mean, variation and the
 $t$ for a given harmonic can be calculated  from the
positions using Eqs. (2-3). But
what is the accuracy of this harmonic calculated for finite $N$ 
compared with the ideal case, when one would have $N \rightarrow \infty$?
In other words, is the calculated $t$ a "true" estimator of the
Student $t$? Obviously, we would need to compare 
the measured $t$ obtained for the "infinite" case with the theoretically 
predicted zero mean. 

Several authors have investigated the answers to these and similar questions
(\cite{trumpler}, \cite{kendall},  
\cite{brace}, \cite{peebles}, \cite{press}, \cite{tegmark}).
Note that some conclusions can be deduced for this case
immediately from the well-discussed case of discrete
Fourier transform (cf. Bracewell 1978, Chapt. 14; Press et al. 1992, Chapts.
12-13); in $\alpha$ we have in fact a Fourier decomposition.
The key result of all these studies is the statement that for
the lowest-order harmonics (dipole, $n= 1$; quadrupole, $n=2$; etc...)
the Student $t$ calculated from the observational data 
can be considered as if it were $N \rightarrow \infty$
(\cite{horack}, \cite{tegmark}). Hence, for these
low-order harmonics Student's $t$ test gives the correct conclusion. 
This result also suggests that
the $2N$ independent harmonics, which should be tested, should be the lowest
possible ones with $n \leq (\sqrt{2N} -1)$. (Clearly, up to $n$, we have 
 $1+3+...+ (2n+1) = (n+1)^2$ harmonics.) In addition, one should note
that, even for 
$n > \sqrt{2N}-1$ some harmonics, being expressed by the lower ones,
can also be used with care.
For example, in Tegmark et al. (1996),  $N$ was $ = 1122$ and even $n=65$
was discussed. Nevertheless, in this paper, in order to avoid 
any complications with higher $n$,
we restrict ourselves to $n \leq (\sqrt{2N} -1)$.

A rough estimate of the inaccuracy of 
quantities in Eqs. (2-3) for the given $n, m, N$ may be 
quite simply performed. 
The typical angular scale belonging to $n$ is 
$\simeq \pi/n$ (Peebles 1980, Chapt. 46), and
hence the corresponding solid angle is
$\simeq (\pi/n)^2$ steradians. Clearly, on a
solid angle $\simeq 4\pi/N$ steradian the discreteness is crucial, and hence
for $(\pi/n)^2 \simeq (4\pi/N)$ the inaccuracies can be even of order unity
(they need not be, but they can be for the worst case).
All this suggests that for
$n$ the inaccuracy coming from discreteness should be maximally
$ \simeq (4\pi/N)/(\pi/n)^2 \simeq n^2/N$. Of course, in the inaccuracy
formula $\simeq n^2/N$ there is an uncertainty concerning
the numerical factor itself. (We have roughly taken $4 \simeq \pi$;
in Eq. (3) $\cos^2 n\alpha$ may be expressed by $\cos 2n\alpha$, and
hence here one will have generally a four-times larger uncertainty; 
 the typical $\pi/n$ angle belonging to $n$ is a rough estimate; etc.)
Recently, this question has usually been  studied
by numerical simulations (cf. Tegmark et al. 1996, and references therein).
But, classical mathematical treatments are also known
(cf. Boas 1983, Chapt. 16.6; Press et al. 1992, Chapt. 13.4;
Bracewell 1978, Chapt. 14). All these studies suggest that for
$n \ll \sqrt{2N}$ the discreteness leads to negligible inaccuracies,
which is in accordance with our estimate $\simeq n^2/N$. On the other
hand, Tegmark et al. (1996) imply that even for $n^2 \simeq N$ the analysis
based on spherical harmonics can be good. This suggests that the inaccuracy
should be much smaller that $n^2/N$. (Note that, trivially, the dependence on
$m$ need not be mentioned especially. Clearly $m \leq n$, and there are
$2n+1$ harmonics for a given $n$. Then
the inaccuracy for a given $n$ will simply
be the largest inaccuracy among these $2n+1$ harmonics.)
We will not go into the details of this discussion, and
in this paper we will simply assume that the inaccuracy is maximally
$\simeq n^2/N$.
We will also keep in mind the fact that the estimate
of this inaccuracy is never complete.
Therefore, one should be careful with any conclusions derived, for 
instance, when $ n^2 \ll N$ does not apply. To avoid problems 
with higher harmonics, we will carry out a further 
drastic truncation.
We will, ad hoc, consider only $n \leq \sqrt{N}/3$. This means that
instead of the possible $2N$ harmonics, we will study only $\simeq N/9$.
This (together with the avoiding of $m = 0$ terms) means that we will test
only $\simeq (5-6)\%$ of  all allowed independent 
harmonics. 

Concerning the "sign-test" the accuracy seems to be much better than
for  Student's $t$ test. By this we mean to say that
 there is no problem with the correctness
of a conclusion coming from the Bernoulli test. Therefore, the "sign-test"
may be used for the same harmonics for which Student's $t$ test was used.

For the used values of $n$ the positional errors (of sizes $\simeq 1-4$
degrees, \cite{fishman94}) should give no further complications. We will 
consider only such values of $n$, when the typical angular size
belonging to an $n$ ($ \simeq \pi/n$; Peebles 1980, Chapt. 46)
will be much larger than the size of the positional error. 

The detailed discussion presented by Tegmark 
et al. (1996) shows that for such values of 
$n$ no complications should arise from the positional errors.

   \section{Results}

In order to test the isotropy of 2281 GRBs collected
in Current BATSE Gamma-Ray Burst 
Catalog (\cite{meegan}) until the end of  1998,
we calculated their spherical harmonics up to $n=15$.
This maximum $n$ value was chosen
in accordance with the discussion in  Sect.2 , where
we restricted ourselves to $n \leq
\sqrt{N}/3$. Here $N=2281$, and $\sqrt{N}/3 = 15.9$.
In other words, instead of $2\times 2281= 4562$ 
possible independent spherical
harmonics we tested only $(16^2 - 16) = 240$ both
with Student's $t$ test and sign tests.
The results are collected in Table 1,
in which we present all dipole and quadrupole terms (in order to
compare them with Bal\'azs et al. 1998), but for larger harmonics only
the terms, for which either $|t| >1.96$ or $k_{\mathrm{obs}} < 1093$ or 
$k_{\mathrm{obs}} > 1188$.
For $|t| > 1.96$ there is a smaller than 5$\%$ 
probability that there is still
an isotropy; for $|t| > 2.58$ this probability
is smaller than $1\%$ (\cite{trumpler}). For the "sign test" one needs 
$|k_{\mathrm{obs}obs} - N/2| > \sqrt{N}$ to have a $>95.4\%$ significance level 
for anisotropy, where $k_{\mathrm{obs}}$ is the observed
number of GRBs having the given harmonic negative value.
We also present the $m = 0$ cases for comparison as well,
but these values will not be taken as indications
of intrinsic anisotropy. For $m=0$ 
the "sign test" is applied only for odd $n$, of course.

    \begin{table}
    \caption{Results of  Student's $t$ and the sign tests, respectively, of
2281 GRBs from spherical harmonics
up to $n=15$. All dipole + quadrupole terms are presented,
but for higher order harmonics only
the components with either $|t| > 1.96$ or $k_{\mathrm{obs}} < 1093$ or 
$k_{\mathrm{obs}} > 1188$. * (**) means that the test gives a $>95 \%$
($>99\%$) probability of anisotropy. For $m >0$ ($m <0$)
one has the cosine (sine) term in Eq. (1). For $n$ even with
$m = 0$ the sign test is not applied. }
   $$
         \begin{array}{rrrrr}
            \hline
            \noalign{\smallskip}
             n & m & t & k_{\mathrm{obs}}\\
            \noalign{\smallskip}
            \hline
            \noalign{\smallskip}
   1 & 0 & *2.29 & 1098 \\
   1 & 1 & -0.44 & 1155 \\
   1 & -1 & 0.12& 1148 \\
   2 & 0 & **4.26 & ... \\
   2 & 1 & -0.88 &1149\\
   2 & -1 & -0.30 & 1149 \\
   2 & 2 & -0.29 & 1137 \\
   2 & -2 & 0.85 & 1137 \\
   \hline
   4 & 4 & 0.89 & *1092 \\
   5 & 4 & *2.20 & 1094 \\
   9 & -6 & 0.53 & *1092\\
   9 & 8 & **2.67 & **1072 \\
   9 & -8 & *2.44 & 1106 \\ 
   10 & -1 & 1.42 & *1089 \\
   10 & -10 & *-2.14 & 1184 \\
   11 & 8 & **2.88 & 1136 \\
   11 & -8 & *2.03 & 1095 \\
   12 & 0 & *2.31 & ... \\
   12 & -12 & **2.69 & 1094  \\
   13 & 8 & **2.81 & **1072 \\
   13 &9 & *2.18 & 1147 \\
   14& -9 & 0.09 & *1092 \\
   14 & -12 & **3.36 & 1103 \\
   15 & 8 & *2.39 & 1115 \\
   15 & 9 & *2.43 & 1110 \\
   15 & 11 & 0.28 & *1091 \\ 
   \hline
   \end{array}
   $$
   \end{table}

Table 1 shows that there are three $m = 0$ components with $|t| >1.96$
($n=1, 2, 12$).
The biggest value is the $n=2, m=0$ quadrupole term with $t = 4.26$.
This means that there exists an anisotropy with a certainty;
the probability that this quadrupole term is non-zero due to
chance is much smaller than $0.1\%$ probability
($t=3.29$ corresponds to a $0.1\%$ probability). In fact, this is nothing new
(\cite{balazs98}); it again confirms that
 the observed distribution of all GRBs is without doubt
anisotropic. 

We will interpret these three terms
 as the anisotropy that follows {\it exclusively} from the non-uniform
sky exposure function of BATSE instrument.

There are no further dipole and quadrupole anisotropies above $|t| > 1.96$;
the sign test also gives no such anisotropies.
On the other hand, there are 19 $m\neq 0$ harmonics 
above the $95\%$ significance level. (In 12 cases $|t| > 1.96$; in 7 cases
the sign test gives anisotropy above the $95\%$ significance level; for
$n=9, m =8$ and $n=13, m =8$, respectively, both tests give anisotropy.)
In addition, for five harmonics one has $|t| > 2.58$. For two of them
 one has $ k_{\mathrm{obs}} < 1093$ as well. Concerning the $95\%$
significance level, one expects that in $2\times 240/20 =24$ cases, one can
obtain this "by chance", too. This is more than the 19 cases obtained, and
the difference could well be due to chance.
Concerning the $99\%$ significance level, one expects 4.8 cases theoretically,
and we obtain 7 ones. This can also be due to chance.
The dispersion arising from the binomial distribution 
is $\sqrt{4.8\times 0.99}= 2.18$; this is practically identical 
to $(7-4.8) = 2.2$. There is also a $t=3.36$ value,
which should give a higher than $99.9\%$ significance level. This single
value is hardly enough to reject the null hypothesis. 
One expects 0.48 such cases. The dispersion is
0.69; hence one single test may well occur "by chance" as well.
Hence, we conclude that the null-hypothesis of intrinsic isotropy
for all GRBs should {\it not} be rejected.

In Current BATSE Catalog there are 1670 GRBs having 
measurements both for the positions and $T_{90}$.
(The durations are catalogued for
GRBs detected before August 1996.) Among  these, 419 
objects have 
$T_{90} < 2$ s ("short" subclass), 253 objects 
have $2$ s $< T_{90} < 10$ s ("intermediate" subclass),
and 998 objects $T_{90} > 10$ s ("long" subclass). 
For these three subclasses we repeated the above procedure.

   \begin{table}
    \caption{Results of the Student $t$ + sign tests, respectively,
for 419 short GRBs from spherical harmonics up to $n=6$.
The notation is the same as in Table 1;
for $k_{\mathrm{obs}}$ here we needed either $k_{\mathrm{obs}} < 190$ or $k_{\mathrm{obs}} > 231$.}

     \label{short-harmonics}
   $$
         \begin{array}{rrrr}
            \hline
            \noalign{\smallskip}
             n & m & t & k_{\mathrm{obs}}\\
            \noalign{\smallskip}
            \hline
            \noalign{\smallskip}
   1 & 0 & 0.38 & 207\\
   1 & 1 & -0.61 & 211 \\
   1 & -1 & 1.52 & 200 \\
   2 & 0 & **2.69 & ... \\
   2 & 1 & -1.07 &218 \\
   2 & -1 & -0.05 & 207  \\
   2 & 2 & -0.08 & 201 \\
   2 & -2 & 1.74 & 199 \\
   \hline
   3 & 1 & 0.98 & *186 \\
   5 &  -1 & 1.60 & *188 \\
   5 & 2  & 0.78 & *189 \\
   5 & 3 &  *-2.19 & 229 \\
 \hline
   \end{array}
   $$
   \end{table}

The results for the 419 short GRBs having $T_{90} < 2$ s are collected
in Table 2. (Here $N=419$, and therefore we consider only $n \leq 6$.)
Again there is a clear non-zero term for $n=2$, $m=0$
 expected from the BATSE's
sky exposure function. There are no dipole and quadrupole terms
above $|t| > 1.96$. On the other hand, there are 4 further tests
determining non-zero harmonics above the $>95\%$ significance
level. Nevertheless, from the 84 tests performed
 here one expects theoretically that 4.2 tests will be 
above the $95\%$ significance level "by chance". Again,
similarly to all GRBs, the null-hypothesis of intrinsic isotropy of
short GRBs should {\it not} be rejected.

   \begin{table}
    \caption{Results of the Student $t$ + sign tests, respectively,
for 253 intermediate GRBs from spherical harmonics up to $n=5$.
The notation is the same as in Table 1;
for $k_{\mathrm{obs}}$ here we needed either $k_{\mathrm{obs}} < 111$ or $k_{\mathrm{obs}} > 142$.}

     \label{intermediate-harmonics}
   $$
         \begin{array}{rrrr}
            \hline
            \noalign{\smallskip}
             n & m & t & k_{\mathrm{obs}}\\
            \noalign{\smallskip}
            \hline
            \noalign{\smallskip}
   1 & 0 & 1.33 & 116\\
   1 & 1 & -0.83 & 136 \\
   1 & -1 & *-2.24 & 142 \\
   2 & 0 & 1.74 & ... \\
   2 & 1 & 1.17 & 120\\
   2 & -1 & -0.68 & 132  \\
   2 & 2 & -0.49 & 129 \\
   2 & -2 & -0.43 & 132 \\
   \hline
   3 & -1 & **-3.37 & **154 \\
   5 & -1 & *-2.25 & 130 \\
   5 & 1  & 0.20 & *108 \\
   5 & 4 &  *2.25 & 112 \\
 \hline
   \end{array}
   $$
   \end{table}

The results for the 253 intermediate GRBs having $2$ s $ < T_{90} < 10$ s 
are collected
in Table 3. (Here $N=253$, and therefore we consider only $n \leq 5$.)
Interestingly, there are no non-zero $m=0$ terms
 expected from the BATSE sky exposure function. There are 6 tests
determining non-zero harmonics above the $>95\%$ significance
level. From the 60 tests done here one expects theoretically
that 3 tests will be above the $95\%$ significance level "by chance". 
The obtained 6 cases - instead of the expected 3 ones - can just still be
by chance. The dispersion is $\sqrt{3\times 0.95} = 1.69$.  
$6-3 = 3$ is still smaller than $2\times 1.69 = 3.38$. Hence the obtained
6 tests instead of the expected 3 can still be by chance with a
probability larger than $5\%$. Nevertheless, we
find that for the intermediate subclass the null hypothesis of isotropy
can be rejected at the $>95\%$ significance level.
This follows from the large value of $t$ and
$k_{\mathrm{obs}}$ for $n=3$, $m=-1$. Student's $t$ test yields a probability of
zero value for this harmonic of $0.08\%$. From the "sign-test" this value
is even smaller, because - from the Bernoulli distribution - 
the zero value is rejected at
the $(154- 126.5)/7.95 = 3.46 \sigma$ significance level; the
probability of a chance is only $0.06\%$. From the considered
60 tests, we obtain 2 cases giving a $\geq 99.92\%$ significance level;
the expected value is maximally $0.48$. Then the dispersion is $0.7$, and 
$(1-0.48) = 1.52 > 2\times 0.7$. Hence, there is
a $1.52/0.7 = 2.17 \sigma$ - i.e. $97\%$ - probability
that this is not by chance. 
Taking into account this estimate
we conclude that 
this single spherical harmonic alone is enough to reject the
assumption of isotropy at a higher than $95\%$ significance level.
This significance level is minimally $97\%$.

   \begin{table}
    \caption{Results of the Student $t$ and sign tests, respectively,
for 998 long GRBs from spherical harmonics
up to $n=10$. The notation is the same as in Table 1;
for $k_{\mathrm{obs}}$ here we needed either $k_{\mathrm{obs}} < 468$ or $k_{\mathrm{obs}} > 530$.}
     \label{long-harmonics}
   $$
         \begin{array}{rrrrr}
            \hline
            \noalign{\smallskip}
             n & m & t & k_{\mathrm{obs}}\\
            \noalign{\smallskip}
            \hline
            \noalign{\smallskip}
   1 & 0 & 0.73 & 487 \\
   1 & 1 & 0.57 & 494 \\
   1 & -1 & 0.83 & 486 \\
   2 & 0 & 1.88 & ... \\
   2 & 1 & -0.51 & 499 \\
   2 & -1 & 0.81 & 481 \\
   2 & 2 & 0.20 & 488 \\
   2 & -2 & 0.38 & 498 \\
\hline
   4 & 0 & *-2.19 & ...\\
   4 & 1 & -1.82 & *534\\
   4& -1 & *2.18 & 474\\
   4 & -4 & 1.36 & *466\\
   5& 4 & *2.44 & 468 \\
   6& -1 & *2.06 & 498 \\   
   6& 2 & 1.33 & *460 \\      
   6& -4& 0.61 & *466 \\
   6& -5 & *2.60 & *463\\
   6& 6 & *-2.04 & 523 \\   
   7 & 0 & *2.20 & 469\\ 
   7 & -2 & -0.91 & *536\\
   8 & -1 & *2.31 & 492\\
   8 & -5 & *2.04 & 496\\
   8 & -7 & *1.96 & 528\\         
   10 & -1 & *2.47 & *467 \\
 \hline
   \end{array}
   $$
   \end{table}

The results for the 998 GRBs having $T_{90} > 10$ s are collected
in Table 4. (Here $N=998$, and therefore here we consider only $n \leq 10$.)
First, from these results it follows that
there are the $n=4, 7$, $m=0$ anisotropy terms expected from the BATSE
sky exposure function. There are no further dipole and quadrupole terms
sugges\-ting anisotropy on these scales.

Second, there are 16 tests giving 
anisotropies between $95\%$ and $99\%$
significance levels. Third, there are no tests giving
anisotropy above the $99\%$ probability level.
Here, except for the $m=0$ terms, 220 test were done, and from them
theoretically 220/20 = 11 tests should give $>95\%$ probabilities by
"chance". Instead of this, there are 16 such cases, which can still 
be by chance. The dispersion is $\sqrt{0.95\times 11}= 3.23$, and
 $2\times 3.23 = 6.46 > (16-11) = 5$. The probability to obtain 16 cases
instead of 11 by chance is bigger than $5\%$.
Again, similarly to all GRBs and to the short subclass, 
the null-hypothesis of intrinsic isotropy of
long GRBs should {\it not} be rejected.

   \section{Conclusions}

The results of paper may be summarized as follows.

First, the $n=2$, $m=0$ spherical
harmonic shows that there is a clear anisotropy on the significance level
$>99.9\%$ in the distribution
 of all 2281 GRBs. This fact is expected from the
BATSE sky exposure function, and is interpreted as an artificial
"instrumental" anisotropy. 

Second, there is a clear intrinsic
anisotropy of 253 "intermediate" GRBs at the $\geq 97\%$ significance
level due to the $n=3$ $m=-1$ term.

Third, both the 419 short GRBs and the 998 long GRBs, respectively,
and also  all 2281 GRBs do {\it not} exhibit anisotropies on 
statistically high enough significance levels.

All these results are interesting, because the departure from 
intrinsic isotropy just for the new "intermediate" subclass having 
the smallest number of GRBs is surprising. Of course, the significance 
level should still be increased, because the $99\%$ level (or even 
the $99.9\%$) is desirable. However, the $97\%$ 
significance level is already remarkable, and hence surely
should be announced.
The refinement of significance level may follow either from
the methods used in this
paper and in Bal\'azs et al. (1998)
(for example, a better estimation of inaccuracy
will not need the drastic truncation $n <\sqrt{N}/3$), or from wholly
different statistical methods  (\cite{kendall}, \cite{bagoly}).

   \begin{acknowledgements}
We thank the valuable discussions with L.G. Bal\'azs, I. Horv\'ath, D. James,
J. Katgert, P. M\'esz\'aros, D. Vokrouhlick\'y and 
with the referee, S. Sigurdsson. This article was partly supported 
by GAUK grant 36/97, by GA\v{C}R grant 202/98/0522, and by
Domus Hungarica Scientiarium et Artium grant (A.M.).
   \end{acknowledgements}

\end{document}